\documentclass[conference,anonymous]{IEEEtran}
\IEEEoverridecommandlockouts
\usepackage{cite}
\usepackage{amsmath,amssymb,amsfonts}
\usepackage{algorithmic}
\usepackage{graphicx}
\usepackage{textcomp}
\usepackage{xcolor}
\usepackage{url}
\usepackage{enumitem}
\usepackage{threeparttable}
\usepackage{multicol}
\usepackage{multirow}
\usepackage{xspace}
\usepackage{tikz}
\usepackage{tcolorbox}
\usetikzlibrary{patterns}
\usepackage{pgfplots}
\usepackage[skip=0pt,font=footnotesize]{caption}
\usepackage{booktabs}
\def\BibTeX{{\rm B\kern-.05em{\sc i\kern-.025em b}\kern-.08em
    T\kern-.1667em\lower.7ex\hbox{E}\kern-.125emX}}

\newcommand{\ie}{\textit{i.e.,}\xspace}
\newcommand{\eg}{\textit{e.g.,}\xspace}

\newcommand{\etal}{et al.\xspace}

\newcommand*{\thead}[1]{\multicolumn{1}{c}{\bfseries #1}}
    
\begin{document}

\title{Exploring the Evidence-Based SE Beliefs of Generative AI Tools
\thanks{979-8-3315-9147-2/25/\$31.00 ©2025 IEEE}
}

\author{\IEEEauthorblockN{Chris Brown}
\IEEEauthorblockA{\textit{Virginia Tech}\\
dcbrown@vt.edu}
\and
\IEEEauthorblockN{Jason Cusati}
\IEEEauthorblockA{\textit{Virginia Tech} \\
djjay@vt.edu}
}
\maketitle

\begin{abstract}
    \underline{\textit{Background:}} Recent innovations in generative artificial intelligence (AI) have transformed how programmers develop and maintain software. The advanced capabilities of generative AI tools in supporting development tasks have led to a rise in their adoption within software engineering (SE) workflows. However, little is known about how AI tools perceive evidence-based practices supported by empirical SE research. \underline{\textit{Aim:}} To this end, we explore the ``\textit{beliefs}'' of generative AI tools increasingly used to support software development in practice. \underline{\textit{Method:}} We conduct a preliminary evaluation conceptually replicating prior work to investigate 17 evidence-based claims across five generative AI tools. \underline{\textit{Results:}} Our findings demonstrate generative AI tools have ambiguous beliefs regarding research claims and lack credible evidence to support responses. \underline{\textit{Conclusions:}} Based on our results, we provide implications for practitioners integrating generative AI-based systems into development contexts and shed light on future research directions to enhance the reliability and trustworthiness of generative AI---aiming to increase awareness and adoption of evidence-based SE research findings in practice.
\end{abstract}


\section{Introduction}
Recent technological advancements have transformed how software is constructed and maintained. For example, generative artificial intelligence (AI) tools, powered by large language models (LLMs), are increasingly capable of supporting software engineering (SE) tasks, including software design~\cite{ahmad2023towards}, repair~\cite{sobania2023analysis}, implementation~\cite{liu2024your}, and testing~\cite{yuan2023no}. As such, developers are increasingly integrating generative AI into SE workflows. For instance, studies show 84\% of developers are familiar with generative AI tools~\cite{jetbrains} and 76\% are or plan to adopt them~\cite{stackoverflowAI2024}. Moreover, GitHub reports approximately half of developers' code across programming languages is assisted by GitHub Copilot, an 
LLM-based programming assistant~\cite{zhao_copilot}.


To support software and its engineering, empirical SE research seeks to understand and enhance software development. These findings aim to promote empirically-based decision making in SE contexts to enhance software products~\cite{kitchenham2004evidence}. Yet, software engineers largely ignore research findings in practice---finding the results to be irrelevant in industry contexts~\cite{lo2015practitioners,winters2024thoughts}. For example, while research shows automated tools are useful for debugging code and benefiting software projects~\cite{GoogleFixit}, developers often avoid them~\cite{Johnson2013Why} and ignore notifications from these systems~\cite{nasif2019challenges} in practice. 

Generative AI is increasingly used to automate tasks within software development workflows~\cite{li2025unveiling}. However, little is known about the perceptions (or ``\textit{beliefs}'') of these systems regarding empirical SE research claims. As humans and generative AI increasingly work together in SE contexts, understanding these beliefs is crucial. Software influences a growing extent of modern society, and generative AI is increasingly used to support SE~\cite{ozkaya_2023}---impacting the behavior and decision-making of software engineers~\cite{vereschak2021evaluate}. To this end, we aim to answer the following research questions (RQs):
\begin{itemize}[noitemsep]
    \item[] \textbf{RQ1: What do generative AI tools believe about empirical SE research claims?} 
    \item[] \textbf{RQ2: How do generative AI tools obtain evidence to support their beliefs?}
\end{itemize}

Prior work explores the beliefs and behaviors of human software engineers with regard to empirical SE claims~\cite{devanbu2016belief}. We extend this work by conducting a \textit{conceptual replication study}~\cite{crandall2016scientific} to observe the evidence-based beliefs of generative AI tools used in SE tasks. We test the same fundamental ideas examined by Devanbu \etal~\cite{devanbu2016belief}, prompting AI-powered systems instead of human developers to investigate beliefs regarding evidence-based SE research findings. We found generative AI mostly agrees with evidence-based SE claims, but provide ambiguous answers and lack credible sources of evidence. Based on these emerging results, we provide implications for developers utilizing LLM-driven systems to complete development tasks. We also offer research directions to enhance development workflows and leverage generative AI to improve the adoption of evidence-based practices in SE.

\textbf{\em Motivating Example:} Consider the generative AI-based tool Devin\footnote{\url{https://devin.ai/}}---recently announced as ``the first AI software engineer''. Devin is a novel AI system that works with developers to generate code, fix bugs, add new features, and deploy projects in its own development environment~\cite{devin}. In three demonstration videos for Devin, the AI software engineer uses print statements to debug errors in code,\footnote{See: \url{https://www.youtube.com/watch?v=fjHtjT7GO1c} (0:52--1:10); \url{https://www.youtube.com/watch?v=TiXAzn2_Xck} (2:12--2:25); and \url{https://www.youtube.com/watch?v=ReE2dFJn_uY&t=50s} (0:40--0:49)} despite being equipped with its own coding editor and common development tools. While print-statement debugging is widely adopted by programmers in practice~\cite{beller2018dichotomy,min2023debugging}, it is discouraged by research in favor of automated debugging techniques~\cite{lieberman1997debugging}. 

\section{Background}


\subsection{Terminology}\label{sec:terms}

\subsubsection{Generative AI Tools} LLMs are deep learning models trained on large amounts of data~\cite{ozkaya_2023}. In this work, \textit{generative AI tools} are chat-based systems powered by LLMs capable of assisting in coding and SE tasks. For this project, we examine five tools: \textbf{ChatGPT}~\cite{gpt} is a chatbot by OpenAI. We accessed ChatGPT 3.5 through its web interface online; \textbf{GitHub Copilot}~\cite{copilot} is a programming assistant powered by OpenAI's GPT-4 model. We accessed GitHub Copilot Chat via Visual Studio Code;\footnote{\url{https://code.visualstudio.com/}, Version 1.87.1} \textbf{Gemini}~\cite{gemini} is powered by Google AI and trained on information from Google services. We utilized the Basic version of Gemini 1 online; \textbf{Blackbox AI}~\cite{blackbox_ai} claims to be the ``best AI model for code''.\footnote{\url{https://www.blackbox.ai/}} We used Blackbox AI Code Chat version 2.2.10 accessed online; \textbf{Claude}~\cite{claude} is an AI system from Anthropic that supports code generation. We accessed Claude 3 Haiku online.

\subsubsection{Empirical SE Claims}

We define \textit{empirical SE claims} as findings derived through research methods that regard factors related to software development. Researchers offer insights to guide systematic efforts to support data collection and analysis in SE contexts~\cite{storey2020software}. For our study, we derived empirical SE claims from prior work exploring the beliefs of empirical SE findings from the perspective of human software engineers~\cite{devanbu2016belief}. In addition, we add an additional claim related to the use of debugging tools based on our motivating example of Devin using print-statement debugging to find and fix coding errors. The list of claims is presented in Table~\ref{tab:claims}. 

\subsubsection{Generative AI Beliefs}

We define \textit{belief} as an assumption made in natural language responses from systems powered by generative AI. While there are numerous uses of ``belief'' in AI across philosophical and cognitive science domains~\cite{hadley1991many}, we adopt the syntactically defined logic-based ``belief as `what one could rapidly discover''', presented by Hadley as follows:

\begin{quote}
    \small{Agent \textsl{X} believes [sentence] \textsl{S} if and only if \textsl{S} is in \textsl{X}'s belief base, or \textsl{X} could prove \textsl{S}, using the belief base and a logic \textsl{L}, within time \textsl{T} belief base.~\cite[p.~66]{hadley1991many}}
\end{quote}

This definition incorporates the capabilities of modern generative AI tools---which are trained on information provided in a large belief base, but have the ability to generate new knowledge in a fixed amount of time---as opposed to other approaches (\ie belief as ``degrees of confidence''). Further, this definition of belief regarding AI can impact human reasoning~\cite{hadley1991many}. We focus on generative AI beliefs with regard to research-based claims related to software development.

\subsection{Related Work}\label{sec:related}

\subsubsection{Beliefs in SE}

Empirical SE is a subset of SE research based on observed and measured phenomena~\cite{empirical}, in software development contexts~\cite{storey2020software}. Many studies demonstrate empirical SE findings are \textit{not} adopted by practitioners (\ie~\cite{Johnson2013Why}). Passos \etal found developers rely on personal experiences for beliefs~\cite{passos2011analyzing}. Researchers suggest SE can benefit from evidence-based practices~\cite{dyba2005evidence}, and offer steps to promote empirically-based decisions~\cite{kitchenham2004evidence,pfleeger2025evidence}. Prior work also investigates if developers' beliefs vary based on personality~\cite{smith2016beliefs} and impact software quality and productivity~\cite{shrikanth2021assessing}. Devanbu \etal show varied beliefs among programmers when assessing empirical SE claims~\cite{devanbu2016belief}, which forms the basis of this study. We extend this work by investigating the beliefs of generative AI tools---which are increasingly used in conjunction with human programmers in SE contexts.

\subsubsection{Generative AI in SE}

There is emerging research interest on generative AI in SE. Zheng et al. outline 123 research papers exploring LLMs across SE tasks, including code generation, vulnerability detection, code analysis, and Q\&A interactions~\cite{zheng2023towards}. Generative AI can support various tasks in the software development lifecycle, such as refactoring, coding, and design~\cite{white2023chatgpt}. Research also outlines limitations of generative AI in SE. For instance, Wang \etal suggest developers find ChatGPT excels at simple coding tasks, but lacks the ability to support other SE activities~\cite{wang2024rocks}. Studies also show theses systems can be inaccurate in SE contexts~\cite{fan2023large}. We extend these efforts to investigate generative AI tools regarding their ability to support evidence-based SE research claims.

\subsubsection{Generative AI in Research}

Prior work investigates generative AI in research contexts. Auer et al. assess the ability of ChatGPT to answer research-based questions from varying domains~\cite{auer2023sciqa}. In SE contexts, research explores if LLMs can emulate humans in surveys~\cite{steinmacher2024can}, replicate empirical SE research findings~\cite{liang2024gpt4}, and offers guidelines for usage~\cite{trinkenreich2025get}. Prior work uses generative AI to provide evidence-based responses to practical SE-related questions through a GPT-powered chatbot~\cite{researchbot}. We seek to understand how generative AI tools perceive empirical SE research claims.


\section{Methodology}


\subsection{Data Collection}


To observe the beliefs of generative AI tools, we used prompting---a key feature of conversational LLMs where users submit queries and receive generated natural language responses~\cite{zamfirescu2023johnny}. Replicating prior work exploring human software engineers' beliefs, we used zero-shot prompting to prompt each tool regarding empirical SE research claims once without providing examples or clarification~\cite{zero}. 

\begin{table*}[]
    \centering
    \footnotesize
        \caption{Empirical Software Engineering Research Claims}
    \begin{threeparttable}
    \begin{tabular}{p{\linewidth}}
    \hline
    \thead{Claim}    \\ \hline
    \rule{0pt}{2.1ex} \hangindent=1.5em \textbf{C1:} Code quality (defect occurrence) depends on which programming language is used. \\
    \rule{0pt}{2.1ex} \hangindent=1.5em \textbf{C2:} Fixing defects is riskier (more likely to cause future defects) than adding new features. \\
    \rule{0pt}{2.1ex} \hangindent=1.5em \textbf{C3:} Geographically distributed teams produce code whose quality (defect occurrence) is just as good as teams that are not geographically distributed. \\
    \rule{0pt}{2.1ex} \hangindent=1.5em \textbf{C4:} When it comes to producing code with fewer defects specific experience in the project matters more than overall general experience in programming. \\
    \rule{0pt}{2.1ex} \hangindent=1.5em \textbf{C5:} Well commented code has fewer defects. \\
    \rule{0pt}{2.1ex} \hangindent=1.5em \textbf{C6:} Code written in a language with static typing (e.g., C\#) tends to have fewer bugs than code written in a language with dynamic typing (e.g., Python). \\
    \rule{0pt}{2.1ex} \hangindent=1.5em \textbf{C7:} Stronger code ownership (i.e, fewer people owning a module or file) leads to better software quality. \\
    \rule{0pt}{2.1ex} \hangindent=1.5em \textbf{C8:} Merge commits are buggier than other commits. \\
    \rule{0pt}{2.1ex} \hangindent=1.5em \textbf{C9:} Components with more unit tests have fewer customer-found defects. \\
    \rule{0pt}{2.1ex} \hangindent=1.5em \textbf{C10:} More experienced programmers produce code with fewer defects. \\
    \rule{0pt}{2.1ex} \hangindent=1.5em \textbf{C11:} More defects are found in more complex code. \\
    \rule{0pt}{2.1ex} \hangindent=1.5em \textbf{C12:} Factors affecting code quality (defect occurrence) vary from project to project. \\
    \rule{0pt}{2.1ex} \hangindent=1.5em \textbf{C13:} Using asserts improves code quality (reduces defect occurrence). \\
    \rule{0pt}{2.1ex} \hangindent=1.5em \textbf{C14:} The use of static analysis tools improves end user quality (fewer defects are found by users).\\
    \rule{0pt}{2.1ex} \hangindent=1.5em \textbf{C15:} Coding standards help improve software quality.\\
    \rule{0pt}{2.1ex} \hangindent=1.5em \textbf{C16:} Code reviews improve software quality (reduces defect occurrence). \\
    \rule{0pt}{2.1ex} \hangindent=1.5em \textbf{C17*:} Debugging tools improve software quality (reduce defect occurrence) more efficiently than print-statement debugging. \\ \hline
    \end{tabular}
    \begin{tablenotes}
\centering
\item \footnotesize{* Evidence-based SE claims derived from~\cite{devanbu2016belief}, C17 is unique to this research.}
\end{tablenotes} 
\end{threeparttable}
    \label{tab:claims}
\end{table*}

\subsubsection{RQ1: Beliefs}\label{sec:collect-beliefs}

We observed responses from generative AI tools to determine their beliefs. To prompt each tool, the empirical SE research claims (see Table~\ref{tab:claims}) were transformed into questions. For example, ``Code quality (defect occurrence) depends on which programming language is used'' was input as ``Does code quality (defect occurrence) depend on which programming language is used?'' (C1). We initially devised prompts to provide a 5-point Likert response from one (Strongly Disagree) to five (Strongly Agree) for each claim, as done in the original study~\cite{devanbu2016belief}---however, we found most AI tools generated arguments for each rating without selecting a level of agreement. Thus, we transitioned to prompting with yes or no questions---limiting our results. This still provided ambiguous responses, which we incorporate in our analysis. 

\subsubsection{RQ2: Evidence}\label{sec:collect-evidence}
To examine how AI tools evince their beliefs, we analyzed generated responses outlining beliefs (see Section~\ref{sec:collect-beliefs} and Table~\ref{tab:claims}). In addition, we prompted tools to understand how they came up with responses. Following each prompt asking about empirical research claims, we prompted ``How did you arrive at your answer?'' to gain further insight into how generative AI forms the basis of their beliefs.

\subsection{Data Analysis}

To analyze our data, we used a mixed methods approach.

\subsubsection{RQ1: Beliefs}\label{sec:belief-analysis} We used closed coding to analyze generated beliefs, categorizing responses as ``Yes'' (Y), ``No'' (N), ``Ambiguous'' (A), or ``Not Applicable'' (N/A)---indicating whether generative AI tool output agreed with, disagreed with, or provided a noncommittal response (\ie did not make a clear stance) to claims. For example, a \textit{Y} response for C14 signifies generative AI agreed static analysis tools improve code quality. We measured \textit{ambiguous} responses since we do not re-prompt models to obtain definitive answers. Two authors performed the closed coding on all 85 responses---providing individual ratings then meeting to discuss and resolve disagreements. We observed ``fair agreement'' (Cohen's $\kappa = 0.398$)~\cite{landis1977measurement}. 

To further investigate the evidence-based beliefs of generative AI, we calculate the rate of correct responses (\textbf{\checkmark\%}) to observe alignment with research claims and observe which findings are most upheld by AI tools. We also calculate variance ($\sigma^2$) to determine the most controversial beliefs among AI tools, similar to the original study~\cite{devanbu2016belief}. To calculate the variance, the initial categorizations were reclassified as numerical values (Y $\rightarrow$ 1, A $\rightarrow$ 2, and N $\rightarrow$ 3)~\cite[Chapter~3.2]{inference}. Then, variance was calculated using Equation~\ref{ref:eq-variance}.

\begin{figure}[h]
\begin{equation}
\sigma^2 = \frac{1}{n} \sum_{i=1}^{n}(x_i-\bar{x})^2
\label{ref:eq-variance}
\end{equation}
\caption*{Eq. 1. where $n$ represents the total number of samples, $x_i$ corresponds with each value in the dataset, and $\bar{x}$ is the mean of all values in the dataset.}
\end{figure}

\subsubsection{RQ2: Evidence}\label{sec:evidence-analysis} 

To answer RQ2, we inspected the \textit{original response} to claims to determine if generative AI tools provided evidence---\ie resources or data---to support their beliefs. The first author scanned responses for links to and clear mentions of evidence sources. The resulting evidence was categorized into the following types: (1) \textit{research papers} published in academic venues; (2) opinionated \textit{blogs}; (3) \textit{social media} posts; (4) \textit{training resources}, \ie tutorials and documentation; (5) posts on \textit{Q\&A} platforms; (6) \textit{tech reports} for non-academic studies; and (7) AI-generated \textit{code examples}.

We also prompted AI tools to provide evidence for responses (see Section~\ref{sec:collect-evidence}). Two authors used open coding to analyze the \textit{prompted responses}---independently evaluating output to investigate sources of evidence. After developing initial codes, the researchers segmented and grouped responses into categories based on common patterns and recurring themes. We observed four main sources of evidence: research, training data, model reasoning, and previous conversations. Our study materials are available in a public repository.\footnote{\url{https://github.com/code-world-no-blanket/LLM-beliefs}} 

\subsection{Threats to Validity}


\paragraph{External Validity} We study five proprietary generative AI tools trained on varying data from different organizations. However, our results may not generalize to all generative AI tools. We examine limited empirical SE research claims, which were motivated by the original study~\cite{devanbu2016belief}. To enhance generalizability, future work can explore different LLM types (\ie open-source models), more advanced generative AI models (\ie GPT-4), and additional SE research claims.

\paragraph{Construct Validity} Generative AI systems themselves cannot ``believe''---as their output relies on models and training data. Our method to collect beliefs relies on prompting, and we use this term for ``technical convenience, and not as an analysis of `real belief'''~\cite[p.~66]{hadley1991many}. Further inspection and more in-depth knowledge of models is needed to provide better insights on AI perceptions of empirical research claims. 

\paragraph{Internal Validity}  Our study methods have limitations. We use zero-shot prompting to observe initial responses, similar to replicating survey questions to human developers. Yet, LLM-based chatbots can respond differently to the same prompt~\cite{ouyang2023llm}. Techniques such as one- or few-shot prompting and instruction finetuning could impact our findings~\cite{prompt}. For calculating variance, converting categorical data to numerical values can limit our results~\cite{inference}. Our manual qualitative analyses for RQ1 and RQ2 may also incorporate bias. 

\section{Results}\label{sec:results}


\subsection{RQ1: Generative AI-Generated Beliefs}

Our results on the beliefs of generative AI tools are presented in Table~\ref{tab:rq1}. Most responses ($n = 55$, 64.7\%) agree with the evidence-based claims explored in our study. We saw a limited number of negative responses ($n = 9$, 10.6\%) disagreeing with claims, and noted 22 ambiguous responses (25.9\%). We provide further insights on the beliefs and accuracy of generative AI tools below:

\paragraph{Generative AI Beliefs} The \underline{most supported} evidence-based practices were unit testing (C9), assert statements (C13), static analysis tools (C14), coding standards (C15), and code reviews (C16) enhance code quality---to which all generative AI tools agreed. In addition, each AI tool agreed with the claims that more complex code leads to more defects (C11) and quality varies based on project context (C12). The \underline{least supported} claims had no positive responses. For instance, all generative AI tools provided ambiguous responses to statically typed languages having fewer bugs (C6) and stronger code ownership leading to higher quality software (C8). In addition, no AI tool agreed that merge commits are buggier than other types of commits (C8). C8 and C2 (fixing bugs is riskier than adding new features), received the most negative responses ($n = 3$)---despite research evidence supporting these claims.

The \underline{most controversial} beliefs are claims with the highest variance, indicating more disagreement between generative AI tools. We found C2---fixing defects
is more likely to cause defects than adding new features---was the most controversial ($\sigma^2 = 0.96$). While research supports C2~\cite{shihab2012industrial}, we observed three negative and two positive responses. For instance, ChatGPT responded, ``fixing defects is often perceived as less risky than adding new features'' and Gemini responded, ``fixing defects is generally considered less risky because you're dealing with existing code''. Yet, positive responses noted ``defects are often interconnected, and fixing one defect can have a ripple effect, causing other defects to emerge. This can lead to a cascade of fixes, each with its own set of risks'' (Blackbox AI) and `` adding new features typically involves working with a more isolated and well-defined scope, which can be easier to design, implement, and test thoroughly'' (Claude).

 
\paragraph{Generative AI Accuracy} We found Blackbox AI and Claude were the \underline{most accurate} with regard to the empirical SE research findings---providing positive responses for 13 of the 17 claims (76\%). Both models also did not respond negatively to any evidence-based claims, while providing ambiguous responses to four. Alternatively, the \underline{least accurate} models were ChatGPT and GitHub Copilot, agreeing with only nine claims (53\%). Copilot was the most incorrect model, responding negatively to the most evidence-based claims ($n = 4$, 23.5\%), disagreeing that code quality depends on programming language (C1), fixing defects is more likely to cause bugs than adding new features (C2), specific project experience decreases bugs compared to general programming experience (C4), and merge commits are buggier than other types of commits (C8). ChatGPT also responded negatively to C2, C4, and C8, but provided a positive response to C1. ChatGPT, along with Gemini, were the most ambiguous ($n = 5$, 29.4\%)---consistently sharing both reasons to agree with and disagree with empirical research claims without taking a clear stance.

\begin{table*}[]
    \centering
    \footnotesize
    \caption{Generative AI-Generated Belief (RQ1) Results}
    \begin{tabular}{llccccc||rr}\hline
       \textbf{Belief} & \textbf{Summary}   & \textbf{ChatGPT}  & \textbf{GitHub Copilot} & \textbf{Gemini} & \textbf{Blackbox AI} & \textbf{Claude} & \textbf{\checkmark\%} & \textbf{$\sigma^2$}  \\ \hline
        C1 & {\scriptsize Quality depends on language} & Y & N & Y & Y & Y & 0.8 & 0.64 \\ 
        C2 & {\scriptsize Bug fixing is riskier than new features} & N & N & N & Y & Y & 0.4 & 0.96 \\ 
        C3 & {\scriptsize Distributed teams $\approx$ non-distributed} & A & A & Y & Y & Y & 0.6 & 0.24 \\ 
        C4 & {\scriptsize Specific experience $>$ general experience} & N & N & A & Y & Y & 0.4 & 0.80 \\
        C5 & {\scriptsize Well commented code reduces defects} & A & A & A & A & Y & 0.2 & 0.16 \\
        C6  & {\scriptsize Static typed languages have fewer bugs} & A  & A & A & A & A & 0.0 & 0.00 \\ 
        C7  & {\scriptsize Stronger code ownership = better quality}    & A & A & A & A & A & 0.0 & 0.00 \\ 
        C8 & {\scriptsize Merge commits buggier than others}     & N & N & N & A & A  & 0.0 & 0.24 \\ 
        C9 & {\scriptsize More unit testing = fewer defects}     &  Y & Y & Y & Y & Y & 1.0 & 0.00 \\ 
        C10 &  {\scriptsize Experienced programmers = fewer defects}  & A & Y & Y & Y & A & 0.6 & 0.24 \\ 
        C11 & {\scriptsize More complex code = more defects}    & Y & Y & Y & Y & Y & 1.0 & 0.00 \\ 
        C12 & {\scriptsize Quality varies between projects}     & Y & Y & Y & Y & Y & 1.0 & 0.00 \\ 
        C13 & {\scriptsize Assert statements improve code quality}    & Y  & Y & Y & Y & Y & 1.0 & 0.00 \\ 
        C14 & {\scriptsize Static analysis improves end user quality} & Y & Y & Y & Y & Y & 1.0 & 0.00 \\ 
        C15 & {\scriptsize Coding standards improve quality}   & Y & Y & Y & Y & Y & 1.0 & 0.00 \\ 
        C16 & {\scriptsize Code reviews improve quality}   & Y & Y & Y & Y & Y & 1.0 & 0.00 \\ 
        C17 & {\scriptsize Debugging tools improve quality}     &  Y & Y & A & Y & Y & 0.8 & 0.16 \\ \hline \hline
        \textbf{\checkmark\%}  & {\scriptsize [Generative AI tool correctness]} & 0.53 & 0.53 & 0.59 & 0.76 & 0.76 & --- & --- \\
        \textbf{$\sigma^2$}  & {\scriptsize [Generative AI tool variance]} & 0.58 & 0.68 & 0.48 & 0.18 & 0.18 & --- & --- \\ \bottomrule
        
    \end{tabular}
    \label{tab:rq1}
\end{table*}

\begin{tcolorbox} [width=\linewidth, colback=gray!3!white, top=1pt, bottom=1pt, left=2pt, right=2pt,label=RQ1]
\textbf{Finding 1:} Generative AI tools agree with most evidence-based claims, but many responses are ambiguous. 

\end{tcolorbox}

\subsection{RQ2: Generative AI-Generated Belief Evidence}\label{sec:res-rq2}

We inspected original responses and prompted AI tools to determine the evidence systems used to arrive at their beliefs.

\paragraph{Original Responses} Most responses ($n = 64$, 75.2\%) lacked clear evidence to support beliefs. We observed 21 instances across three generative AI tools: Claude ($n = 2$), Gemini ($n = 2$), and Blackbox AI ($n = 17$). For Claude and Gemini, we found generic references---\ie ``research has shown...'' (C1, Claude) and ``studies on this topic...'' (C6, Gemini), without specific references or details. Additionally,  one Gemini response included ``Research suggests...[1]'' (C3), without a citation for ``[1]''. Blackbox AI provided the most evidence---totaling 162 sources. Figure~\ref{fig:types} presents the types of evidence observed. Most sources were links to SE-related blog posts ($n = 46$, 28.4\%) and posts on Q\&A websites ($n = 37$, 22.8\%) (\ie Quora, SE Stack Exchange, and StackOverflow). 

We observed 31 (19.1\%) mentions of academic research representing evidence to support SE claims for Blackbox AI. However, upon further inspection we observed multiple issues. For example, there were four instances of the original paper we base our study on~\cite{devanbu2016belief}. We also observed one broken link. Further, we found duplicated references for the same paper. For example, ``On the Impact of Programming Languages on Code Quality''~\cite{berger} was correctly used as evidence for C1---but Blackbox AI provided four links to this paper on HackerNews, arxiv, an author's personal website, and another hosting site. We also observed instances where Blackbox AI summarized specific research studies (\ie ``A study by Microsoft Research found that the bug density of C\# code was similar to that of Python code'' for C6) without citing a specific paper. 



\begin{figure}
    \scalebox{.75}{
    \begin{tikzpicture}  
  
\begin{axis}[
    xbar,
    ytick             = data,
    symbolic y coords = {Tech Report, Code Example, Social Media, Training Resource, Research Paper, Q\&A,  Blog},
    nodes near coords,
  ]
\addplot coordinates { (3,Tech Report) (3,Code Example) (37,Q\&A) (31,Research Paper) (26,Training Resource) (16,Social Media) (46,Blog) }; 
  
\end{axis}  
\end{tikzpicture}
  }
    \caption{Types of Evidence (RQ2) Provided in Generative AI Belief Responses}
    \label{fig:types}
\end{figure}

\begin{table}[t]
\caption{Types of Evidence (RQ2) Provided in Prompted Responses}
    \centering
    \scriptsize
    \begin{tabular}{|l|r|r|r|r|}
    \hline
         & \textbf{Research} & \textbf{Reasoning} & \textbf{Training Data} & \textbf{Prior Response} \\ \hline
        ChatGPT & 13 & 4 & 16 & 0 \\ \hline
        Copilot & 0 & 0 & 17 & 0 \\ \hline
        Gemini & 16 & 15 & 0 & 1 \\ \hline
        Blackbox AI & 17 & 16 & 0 & 0 \\ \hline
        Claude & 5 & 14 & 15 & 0 \\ \hline
        \textit{Total:} & \textit{51} & \textit{49} & \textit{48} & \textit{1} \\ \hline
        
    \end{tabular}
    \label{tab:rq2}
\end{table}

\paragraph{Prompted Response} We also prompted generative AI tools to understand evidence for beliefs. Our analysis revealed four high-level categories (see Table~\ref{tab:rq2}): research ($n = 51$), reasoning ($n = 49$), training data ($n = 48$), and prior response ($n = 1$). Most tools reported using research to inform their responses. For instance, when describing how it arrived to its response for C1, ChatGPT replied ``My response is informed by research and insights gathered from software engineering literature...Studies and articles often discuss how language choice impacts software quality, citing examples and empirical data''. Further, in the prompted response for C3 Gemini incorrectly reports ``I included a reference to a relevant study to support the claim that distributed teams can achieve high code quality''. We also observed instances of AI tools using their own reasoning (\ie ``I reached this answer by considering several factors...'' (Gemini) and ``I recognized the widespread adoption of coding standards in the software industry, which suggests that they are seen as an effective way to improve software quality'' (Blackbox AI)) and training data (\ie ``I generate responses based on patterns and information in the data I was trained on'' (Copilot)) to support responses. Every prompted response from GitHub Copilot indicated it generates output based on the data it was trained on.

\begin{tcolorbox} [width=\linewidth, colback=gray!3!white, top=1pt, bottom=1pt, left=2pt, right=2pt,label=RQ2]
\textbf{Finding 2:} Most generative AI tools lack evidence for SE claims. Only Blackbox AI provided clear evidence, predominantly from blogs and Q\&A posts. 
\end{tcolorbox}

\section{Implications}\label{ref:discussion}

Generative AI tools are increasingly used to support SE tasks. However, our results show these systems lack evidence-based beliefs---potentially inhibiting their utility for practical software development. Based on these findings, we provide implications for developers integrating generative AI tools and LLMs into development workflows.


\paragraph{\bf\em Generative AI is mostly positive, but can be ambiguous on evidence-based practices} We observed the majority of responses from generative AI tools agreed with SE research claims (see Table~\ref{tab:rq1}). However, developers utilizing these systems should note they can be noncommittal on important concepts. For instance, four generative AI tools provided ambiguous responses on the claim well-commented code has fewer defects (C5), providing responses such as ``this correlation is not absolute'' (ChatGPT). Additional measures should be taken to mitigate ambiguity from AI tools. For example, prior work suggests refining prompts and providing feedback or examples (\ie to incorporate best practices) can enhance ChatGPT in SE contexts~\cite{white2023chatgpt}.

\paragraph{\bf\em AI-generated responses lack credible evidence}\label{sec:imply-rq2} While most generative AI tools claimed to use research, we found a lack of evidence to support their responses. AI tools also frequently used their own reasoning and training data (see Table~\ref{tab:rq2}), which are unknown to users and can be misleading~\cite{blackbox}. Only Blackbox AI consistently provided evidence to support claims made by the generative AI model---but, these were largely from non-scientific sources (see Figure~\ref{fig:types}). We also observed several complications such as duplicated references and missing citations. The evidence provided by Blackbox AI could contribute to its accuracy in beliefs, as one of the most accurate AI tools. However, the overall lack of evidence forming generative AI tool beliefs should caution practitioners using their output to support development tasks.

\paragraph{\bf\em Generative AI beliefs are similar to humans} We conducted a conceptual replication of previous work exploring the beliefs of software engineers at Microsoft~\cite{devanbu2016belief}, substituting generative AI tools for humans. We do not directly compare our results due to different rating scales (Y/N/A vs. 5-point scale). However, at a glance we found similar trends in beliefs. For instance, the four least controversial human beliefs---on code reviews (C16), coding standards (C15), static analysis tools (C14), and assert statements (C13)---were unanimously agreed on by the AI tools observed in our study. Likewise, three of the most controversial human claims---code quality depends on language (C1), fixing defects is riskier than adding new features (C2), and specific project experience matters more than general experience (C4)---were the most varied among AI tools. This motivates the need for future research to enhance evidence-based beliefs and practices for generative AI-based SE tools and human developers.


\section{Research Directions}\label{ref:future-work}


Generative AI tools will continue to innovate software development. As software becomes more complex and widespread, AI adoption is increasing among developers~\cite{stackoverflowAI2024}. Generative AI is also increasingly integrated into a growing number of SE-focused contexts (\ie IDEs,\footnote{\eg \url{https://www.jetbrains.com/ai/}} Q\&A platforms,\footnote{\eg \url{https://stackoverflow.blog/2023/07/27/announcing-overflowai/}} etc.). Thus, promoting evidence-based SE practices is increasingly critical. Based on our findings, we provide avenues for future research to increase the adoption of evidence-based practices and beliefs among generative AI tools and the software developers leveraging them to complete SE tasks.

\paragraph{\bf\em Evidence-Based Generative AI} Our results show AI lacks credible evidence to support claims. To mitigate this, future work can explore training generative AI on evidence-based data. AI-based chatbots have been implemented to provide responses backed by scientific evidence to prompts in various domains. For example, Med-PaLM\footnote{\url{https://sites.research.google/med-palm/}} powers a live Q\&A system for medical inquiries about liver cancer~\cite{qian2024liver}. Specialized models have also been developed for technical fields, such as cryptocurrency.\footnote{\url{https://app.chaingpt.org/}} To promote evidence-based SE, future generative AI tools could incorporate empirical SE research findings in their training data to provide responses backed by research. In addition, future work can highlight code bases shown to be high-quality and developed using evidence-based tools and practices to incorporate in training processes. This could increase the reliability of AI, ideally leading to evidence-based output generated by models. 

\paragraph{\bf\em Evidence-Based AI Development Workflows} In our study, we observed generative AI tools can be ambivalent on development practices supported by empirical SE research. Future work can investigate techniques to enhance AI adoption in SE. For instance, a \textit{layered workflow architecture} can apply evidence-based practices on AI-generated code before the output is presented to users. In this context, the layered workflow can incorporate multiple levels of predefined work steps to help organize, control, and execute activities~\cite{kim2001workflow}. A potential example could be: 

\begin{enumerate}
    \item A user prompts a generative AI tool to generate code;
    \item the AI tool generates code based on the prompt;
    \item the output is directed to a different layer (\ie model or system) which applies evidence-based practices on the generated code (\eg C14, static analysis tools); and 
    \item the generated code and resulting static analysis report are provided to the user to inspect---or, AI-based automated program repair capabilities can be leveraged to fix reported bugs before presenting generated code to users.
\end{enumerate}

Prior work uses layers to improve generative AI integration in SE, such as security within AI/ML pipelines~\cite{camacho2024unlocking}. Recent tools, such as AutoGen,\footnote{\url{https://www.microsoft.com/en-us/research/project/autogen/}} also provide multi-agent support for customized workflows. Future work can explore infrastructure to augment generative AI-based development workflows by incorporating research-based tools and practices. Studies can also investigate the effects of comparable approaches on code quality and developer behavior.

\paragraph{\bf\em Generative AI for Translating Research Evidence} Empirical SE research aims to provide evidence supporting effective software development~\cite{kitchenham2004evidence}. However, translating research findings into practice is challenging~\cite{winters2024thoughts}. Horvitz \etal suggest one of the top priorities for the future of AI is to power resources to bridge the academia-industry gap~\cite{horvitznow}. Thus, future work can explore using generative AI to summarize and distribute SE research findings to practitioners. While our findings show zero-shot prompting regarding evidence-based practices often provides ambiguous and unsupported responses, prior work suggests LLMs are effective for summarizing natural language text and code~\cite{ahmed2022few}. Previous attempts to convey human summaries of research findings have shown to be ineffective~\cite{wilson2024will}. However, future work can explore the potential of AI-generated summaries and the role they can play in supporting the behavior of software practitioners.

\section{Conclusion}

To support software development, empirical research offers a variety of tools and practices to enhance software development. Recently, generative AI tools have transformed the development and maintenance of software---receiving rapid adoption for their ability to automatically support a wide variety of SE-related tasks. However, little is known about these LLM-driven systems with regard to evidence-based practices purported by empirical SE research. We analyze the beliefs of popular generative AI tools used for development tasks through a conceptual replication of prior work~\cite{devanbu2016belief}, and explore the evidence for these beliefs. Our findings show generative AI tools often do not provide a clear stance on SE practices, such as the benefits of well-commented code, and lack credible evidence. Based on these results, we provide implications and future research directions to enhance evidence-based decision making with generative AI in SE contexts.

\bibliographystyle{ieeetr}
\bibliography{main}

\end{document}